\documentclass[10pt,letterpaper]{article}
\usepackage{opex3}
\usepackage{amsmath}
\usepackage{cite}

\begin{document}

\title{Pulsed Sagnac polarization-entangled photon source with a PPKTP crystal at telecom wavelength}

\author{Rui-Bo Jin,$^{1,*}$ Ryosuke Shimizu,$^2$ Kentaro Wakui,$^1$   Mikio Fujiwara,$^1$    Taro Yamashita,$^3$   Shigehito Miki$^3$  Hirotaka Terai,$^3$   Zhen Wang,$^{3, 4}$  and Masahide Sasaki$^1$}
\address{$^1$Advanced ICT Research Institute, National Institute of Information and Communications Technology (NICT), 4-2-1 Nukui-Kitamachi, Koganei, Tokyo 184-8795, Japan\\
         $^2$Center for Frontier Science and Engineering, University of Electro-Communications (UEC), Tokyo 182-8585, Japan\\
         $^3$Advanced ICT Research Institute, National Institute of Information and Communications Technology (NICT),  588-2 Iwaoka, Kobe 651-2492, Japan\\
         $^4$Shanghai Institute of Microsystem and Information Technology, Chinese Academy of Sciences (CAS), 865 Changning Road, Shanghai 200050, China}
\email{$^*$ruibo@nict.go.jp}


\begin{abstract}
We demonstrate pulsed polarization-entangled photons generated from a periodically poled $\mathrm{KTiOPO_4}$ (PPKTP) crystal in a Sagnac interferometer configuration at telecom wavelength.
Since the  group-velocity-matching (GVM) condition is satisfied, the intrinsic spectral purity of the photons is much higher than in the previous scheme at around 800 nm wavelength.
The combination of a Sagnac interferometer and the GVM-PPKTP crystal  makes our entangled source compact, stable, highly entangled, spectrally pure and ultra-bright.
The photons were detected by two superconducting nanowire single photon detectors (SNSPDs) with detection efficiencies of 70\% and 68\%  at dark counts of less than 1 kcps.
We achieved  fidelities of  0.981 $\pm$ 0.0002 for $\left| {\psi ^ -  } \right\rangle$ and 0.980 $\pm$ 0.001 for $\left| {\psi ^ +  } \right\rangle$ respectively.
%
%
This GVM-PPKTP-Sagnac scheme is directly applicable to quantum communication experiments at telecom wavelength, especially in free space.
\end{abstract}


%
%
\bibliographystyle{osajnl}


\section{Introduction}
Polarization is an important degree of freedom for photons.
Polarization-entangled photons are fundamental quantum resources in quantum information processing for many applications, such as quantum key distribution \cite{Poppe2004}, photon amplifiers \cite{Gisin2010}, quantum teleportation \cite{Yin2012} and quantum computation \cite{Kok2007}.
The widely used technique for generating the polarization-entangled state is based on a spontaneous parametric downconversion (SPDC) process, which can be arranged in various configurations and with different crystals \cite{ Edamatsu2007,Kwiat1995,Shi2004,Altepeter2005, Li2006, Hentschel2009,Jin2013PRA2,Steinlechner2013}.
Recently, polarization-entangled photons from a periodically poled $\mathrm{KTiOPO_4}$ (PPKTP) crystal in a Sagnac interferometer configuration has become a hot topic, since this source has the merits of  compactness, stability and high brightness.
Kim \emph{et al}, demonstrated the first  entangled photon source with a PPKTP crystal in a Sagnac-loop with a continuous-wave (cw) pump laser at 405 nm in 2006 \cite{Kim2006, Wong2006}.
Then, Fedrizzi \emph{et al}, presented an optimized  scheme with a cw pump laser in 2007 \cite{Fedrizzi2007}.
However, the cw pumped SPDC source cannot provide any timing information about when the photon pair is generated, which is important for  applications such  as synchronization with system clocks in quantum communication systems.
Therefore, Kuzucu and Wong developed a pulsed Sagnac polarization-entangled source at 780 nm in 2008 \cite{Kuzucu2008}.
In 2012, Predojevi\'{c} \emph{et al}, investigated the phase property of this system \cite{Predojevic2012}.
Now this PPKTP-Sagnac scheme has become a common tool for the generation of polarization-entangled state and has applied in many experiments \cite{Fedrizzi2009, Prevedel2011, Vermeyden2013, Ramelow2013, Giustina2013, Cao2013}.

However, in all the previous experiments \cite{Kim2006,  Wong2006, Fedrizzi2007, Kuzucu2008,  Predojevic2012, Fedrizzi2009, Prevedel2011, Vermeyden2013, Ramelow2013, Giustina2013, Cao2013},  the entangled photons were generated at around 800 nm wavelength.
In this work, we expand such pulsed PPKTP-Sagnac scheme into a group-velocity matched (GVM) regime  and demonstrate a GVM-PPKTP-Sagnac scheme at the telecom wavelength.
The concept of GVM in SPDC was introduced by Grice and Walmsley \cite{Grice1997}, and by Keller and Rubin \cite{Keller1997} in 1997.
PPKTP crystal with GVM condition was first experimentally investigated by K{\"o}nig and Wong \cite{Konig2004} for second-harmonic generation  in 2004.
The GVM condition in SPDC has been experimentally realized in KDP crystal at 830 nm wavelength \cite{Mosley2008, Jin2011}, PPKTP crystal at around 1584 nm wavelength \cite{Jin2013OE, Evans2010, Eckstein2011}, and BBO crystal at 1550 nm wavelength \cite{Lutz2013OL, Lutz2013}.
The GVM condition in  spontaneous four-wave mixing has also been demonstrated  in optical fibers \cite{Smith2009}.
In the  case of PPKTP crystal, the GVM condition means $2V^{-1}_{g,p}=V^{-1}_{g,s}+V^{-1}_{g,i}$,
where $V^{-1}_{g,\mu} (\mu=p,s,i)$ is the inverse of the group velocity  $V_{g,\mu}$ for the pump $p$, the signal $s$,  and the idler $i$.

\begin{figure}[tbp]
\centering\includegraphics[width=12cm]{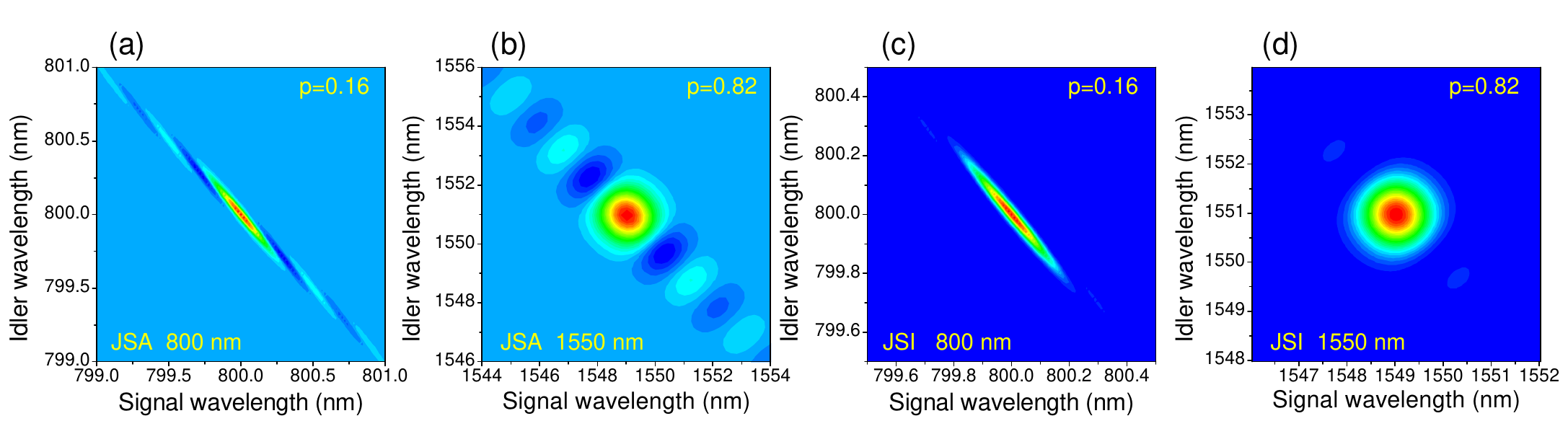}
\caption{ Typical joint spectral amplitude (JSA, a, b) and  joint spectral intensity (JSI, c, d) of the down-converted photons from a PPKTP crystal at 800 nm (a, c) and 1550 nm (b, d), with corresponding maximal spectral purities (p) of  0.16 and 0.82, respectively.
In this simulation, we fixed the crystal lengths at 30 mm, and scanned the full width at half maximum (FWHM) of the pump so as to obtain the maximal purities.
For (a, c), with a pump laser at 400 nm, the  maximal purity was achieved at 0.16 with an FWHM of 0.014 nm (16.8 ps), and  for (b, d) with a pump laser at 775 nm, the maximal purity was 0.82 with an FWHM of  0.4 nm (2.3 ps).
(a, c) were calculated with the Sellmeier equations from \cite{Fan1987} for y direction and \cite{Fradkin1999} for z direction.
(b, d) were calculated with the Sellmeier equations from \cite{Konig2004} for y direction and \cite{Fradkin1999} for z direction.
The spectra of the signal and idler photons in (b, d) have a  Gaussian shape with a bandwidth of around 1.2 nm.
See \cite{Jin2013OE} for more details of the simulations (b, d).
 } \label{2JSI}
\end{figure}

With the GVM condition, the PPKTP crystal may have a high spectral purity at telecom wavelengths \cite{Jin2013OE}.
The spectral purity, a parameter describing the degree of spectral uncorrelation between the signal and idler photons,  is defined as $p = Tr(\hat \rho _s^2 ) = Tr(\hat \rho _i^2)$,
where $\hat \rho_{s(i)} $ is the reduced density operator of the signal ($s$) or idler ($i$), and $Tr$ represents the partial trace.
The spectral purity is calculated by applying Schmidt decomposition on the join spectral amplitude of the signal and idler photons.
See \cite{Jin2013OE} for more details about spectral purity.
See \cite{Evans2010, Eckstein2011,  Gerrits2011, Jin2013PRA, Bruno2014} for the experimentally measured joint spectral intensities of the GVM-PPKTP crystal by several different groups.
Here, we compare the maximal intrinsic spectral purity of PPKTP crystal at around 800 nm and 1550 nm in Fig. \ref{2JSI}.
The maximal intrinsic spectral purity of 0.16 at the 800 nm range is much lower than that at the 1550 nm range, 0.82.
%
%
%
%
%
%

Spectral purity is of paramount importance for experiments with multi-entangled-source.
%
For example, in the entanglement swapping \cite{Scherer2011} or  multi-photon entangled state generation experiments \cite{Huang2011, Yao2012}, the spectral purity of each source must be highly pure to achieve high interference visibilities \cite{Pan2012}.
At 800 nm, to improve the  purity from 0.16 to unity, we need to adopt very narrow bandpass filters to improve the purity, and the brightness will be largely decreased.
However, only coarse bandpass filters can improve the purity from 0.82 to near unity at telecom wavelength range \cite{Jin2013OE}.
Therefore, in principle, the GVM-PPKTP-Sagnac photon source at telecom wavelengths might be much brighter than the PPKTP-Sagnac scheme at the 800 nm range for  multi-entangled-source applications.

Besides the high spectral purity,  another important merit of our GVM-PPKTP-Sagnac scheme is that such high-quality polarization-entangled photons at telecom wavelengths  are suitable for long-haul transmission using low-loss optical fibers.
This establishes the basis for many quantum info-communication applications at telecom wavelength.

Furthermore, low-efficiency photon detectors were obstacles for the telecom-band experiments, but the rapid development in superconducting nanowire single photon detector (SNSPD) technologies has overcome this disadvantage.
In this paper, we demonstrate an ultra-bright polarization entangled photon source with high spectral purity, and detected  by state-of-the-art SNSPDs developed by our group.

\section{Experiment  }

\begin{figure}[htbp]
\centering\includegraphics[width=11 cm]{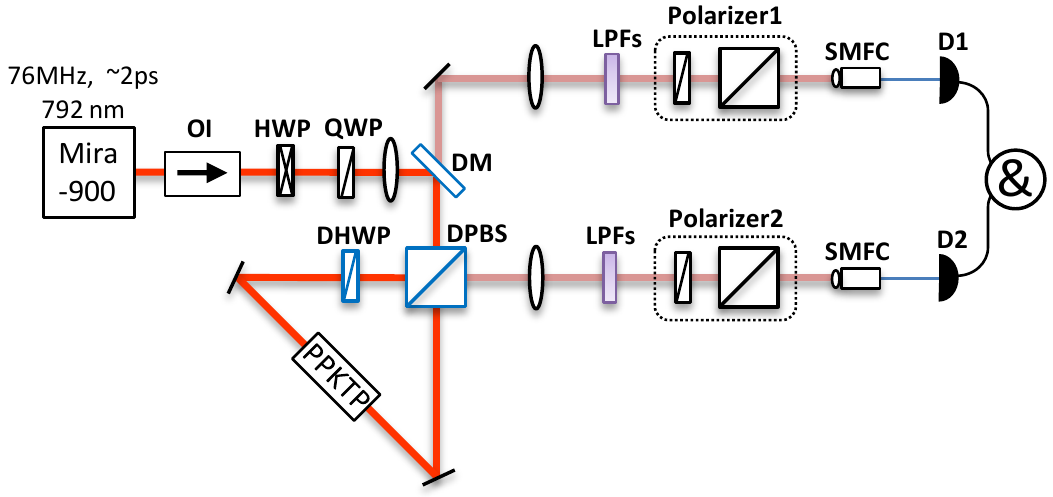}
\caption{ The experimental setup. Picosecond laser pulses (76 MHz, 792 nm, temporal duration $\sim$ 2 ps) from a mode-locked Titanium sapphire laser (Mira900, Coherent Inc.) passed through an optical isolator (OI), a half-wave plate (HWP) and a quarter-wave plate (QWP). Then the pulses were focused by a $f = 200$ mm lens (beam waist $\sim$ 45 $\mu$m), reflected by a dichroic mirror (DM: DMLP1180, Thorlabs) and guided into a Sagnac-loop. The Sagnac-loop consisted of a dual-wavelength polarization beam splitter (DPBS, extinction ratio $=$ 200 : 1, Union Optics), a dual-wavelength HWP (DHWP, for both 792 nm and 1584 nm, Union Optics), and a 30-mm-long PPKTP crystal with a polling period of 46.1 $\mu$m for a type-II collinear group-velocity-matched SPDC. The temperature of the PPKTP was  maintained at $32.5\,^{\circ}\mathrm{C}$  to achieve a degenerate wavelength at 1584 nm.  The PPKTP crystal was pumped by clockwise (CW) and counterclockwise (CCW) laser pulses at the same time. The DHWP is set at 45 degree to make the CCW pump horizontally polarized. The down-converted photons, i.e., the signal and idler,  were collimated by another two $f = 200$ mm lenses, filtered by longpass filters (LPFs) and then coupled into single-mode fibers by two couplers (SMFC). Finally, all the collected photons were sent to two superconducting nanowire single-photon detectors (SNSPDs), which were connected to a coincidence counter (\&). To test the polarization correlation, we inserted two sets of Polarizers (HWP+PBS) before SMFCs. To carry out quantum  state tomography, we replaced the combination of HWP+PBS with that of HWP+QWP+PBS. Since the SNSPDs were polarization dependent, the photons input into the SNSPD were adjusted by fiber-polarization controllers (not shown). The overall efficiency was estimated as 0.10, including the detectors' average efficiency of 0.69, the  SMFCs' average collection efficiency   of 0.23 and  the whole optics' transmission efficiency of  0.64.  } \label{setup}
\end{figure}

The experimental setup is shown in Fig. \ref{setup}.
In this Sagnac interferometer configuration, the pump beam is split into two, the clockwise (CW) pump and the counterclockwise (CCW) pump.
Both the CW and  CCW pump beams are  in opposite directions,  but follow the same path, therefore this scheme is  robust against the optical path changes and can keep phase ultra stable.
Another important feature of this configuration is that the temporal walk-off between the signal (with higher group velocity) and  idler (with lower group velocity)  can be automatically cancelled out, since the signal (idler) generated by CW pump propagates along with the signal (idler) generated by the CCW pump.
A fine alignment of the Sagnac loop is not easy. Therefore, we make a mathematical simulation  \cite{JinMMA} to simulate the beam propagations in a triangle shape Sagnac-loop.
From this simulation, we can learn that the output beams are always in parallel, but never cross.
In order to achieve a completely collinear configuration for both the CW and CCW pump beams, the residual pump beams must overlap with the input laser.

Our superconducting nanowire single photon detectors (SNSPDs)  are fabricated with 5-9 nm thick and 80-100 nm wide niobium nitride (NbN) or niobium titanium nitride (NbTiN)  meander nanowires on thermally oxidized silicon substrates \cite{Miki2013, Yamashita2013}.
The nanowire covers an area of 15 $\mu$m  $\times$ 15 $\mu$m. The SNSPDs are installed in a Gifford-McMahon cryocooler system and are cooled to 2.1 Kelvin.
The maximum system detection efficiency (SDE) is 79\% with a dark count rate (DCR) of 2 kcps.
The measured timing jitter and  dead time (recovery time) were  68 ps \cite{Miki2013} and 40 ns \cite{Miki2007}.
In this experiment, the SDEs of the two SNSPDs were set at 70 \% and 68\%, corresponding to DCRs of less than 1 kcps.
In our previous experiment, we   achieved coincidence counts of 400 kcps (1.17 Mcps) at a pump power of 100 mW (400 mW) with our PPKTP crystal and  SNSPDs \cite{Jin2013SNSPD}.

The output state of this scheme is
\begin{equation}\label{equ1}
\left| \Psi  \right\rangle  \propto  \left| H \right\rangle \left| V \right\rangle  + e^{i\phi } \beta \left| V \right\rangle \left| H \right\rangle,
\end{equation}
where $ \phi $ is the relative phase between the two paths in the CW and CCW directions, and $\beta$ is the ratio of the two pumps  \cite{Kim2006}.
By rotating the angle of QWP and HWP, we  change $ \phi $ and $\beta$.
By slightly moving the position of PPKTP, we can finely adjust the relative phase $ \phi $  by changing the Gouy phase \cite{Predojevic2012, Hamel2010}.
We can easily exchange the state between  $ \left| {\psi ^ -  } \right\rangle  = \frac{1}{{\sqrt 2 }}(\left| {HV} \right\rangle  - \left| {VH} \right\rangle )$  and $\left| {\psi ^ +  } \right\rangle  = \frac{1}{{\sqrt 2 }}(\left| {HV} \right\rangle {\rm{ + }}\left| {VH} \right\rangle )$  by rotating the angle of  QWP and HWP.

\begin{figure}[tbp]
\centering\includegraphics[width=8cm]{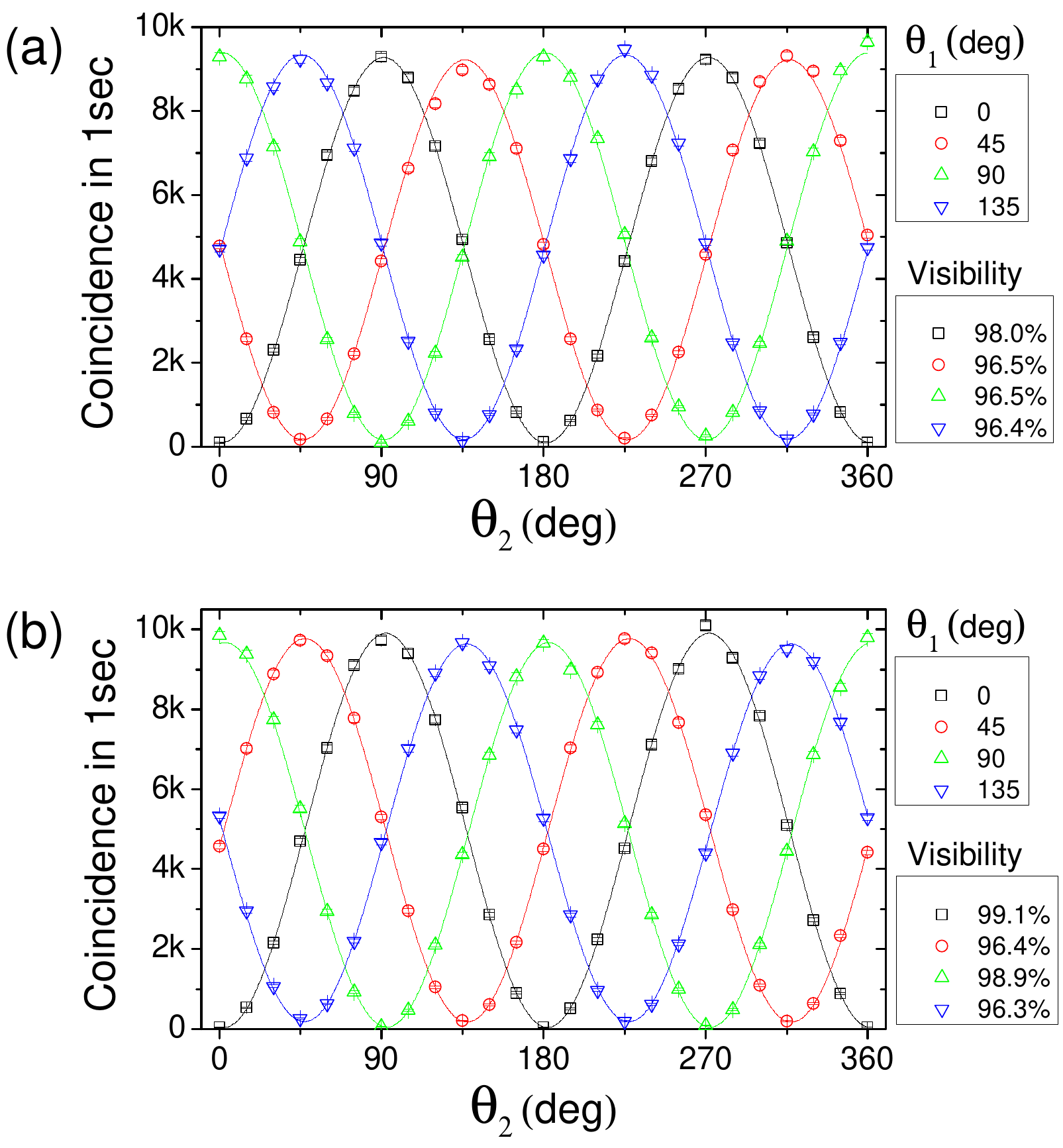}
\caption{ Two-fold coincidence counts in one second as a function of the two polarizers, with a pump power of 10 mW. (a) for $\left| {\psi ^ - } \right\rangle$ state, (b) for $\left| {\psi ^ + } \right\rangle$ state.  The background counts have been subtracted. The error bars were added by assuming Poissonian statistics of these coincidence counts.
  } \label{pattern}
\end{figure}

We set the pump power at 10 mW and carried out a polarization correlation measurement by recording the coincidence counts while changing the angles $\theta_1$ and $\theta_2$  of Polarizer 1 and Polarizer 2, respectively.
The experimental results for some fixed values of $\theta_1$ ( $\theta_1 = 0$, 45, 90 and 135 degrees) are shown in Fig. \ref{pattern}.
For $\left| {\psi ^ -  } \right\rangle$ state in  Fig. \ref{pattern}(a), after  background subtraction, the visibilities were 98.0\% , 96.5\% , 96.5\% , 96.4\%  for Polarizer 1 at 0, 45, 90 and 135 degrees, respectively.
Before background subtraction, the visibilities were 96.7\%, 95.3\%, 95.5\%  and  95.2\%,  respectively.
For $\left| {\psi ^ +  } \right\rangle$ state in  Fig. \ref{pattern}(b), the background subtracted visibilities were 99.1\% , 96.4\% , 98.9\% , 96.3\%  for Polarizer 1 equaled to 0, 45, 90 and 135 degree, respectively.
Before background subtraction, the visibilities were 97.4\%, 95.2\%, 98.0\% and 95.0\%.
Besides the background counts, other two reasons for the degradation of the visibilities were the imperfect compensation of the phase $\phi$ in Eq.(\,\ref{equ1}), and the low extinction ratio of the DPBS (around 200:1).
The measured maximum coincidence count in  Fig. \ref{pattern} was 10 kcps, which corresponded to a coincidence of 20 kcps without polarizers.

All the fringe visibilities in  Fig. \ref{pattern}  were higher than 96\%, which exceeded 71\%, the bound required to  violate the Bell's inequality (also called Bell-CHSH inequality) \cite{Clauser1978}.
We  measured the Bell parameter $S$, which directly indicated the violation of  Bell's inequality \cite{Clauser1969}.
For $\left| {\psi ^ -  } \right\rangle$ state, the obtained value of $S$ was 2.75 $\pm$ 0.01 (75 $\sigma$) with  1 s  accumulation time for each polarizer set, and  2.75 $\pm$ 0.003 (250 $\sigma$) with 10 s.
Without background subtraction, the raw $S$ was  2.72 $\pm$ 0.01 for 1 s, and   2.72 $\pm$ 0.003 for 10 s.
For $\left| {\psi ^ +  } \right\rangle$ state, the obtained value of $S$ was 2.76 $\pm$ 0.01 (76 $\sigma$) for 1 s  accumulation time for each polarizer set,  2.75 $\pm$ 0.003 (250 $\sigma$) for 10 s, and 2.76 $\pm$ 0.001 (760 $\sigma$) for 100 s.

\begin{figure}[tbp]
\centering\includegraphics[width=10cm]{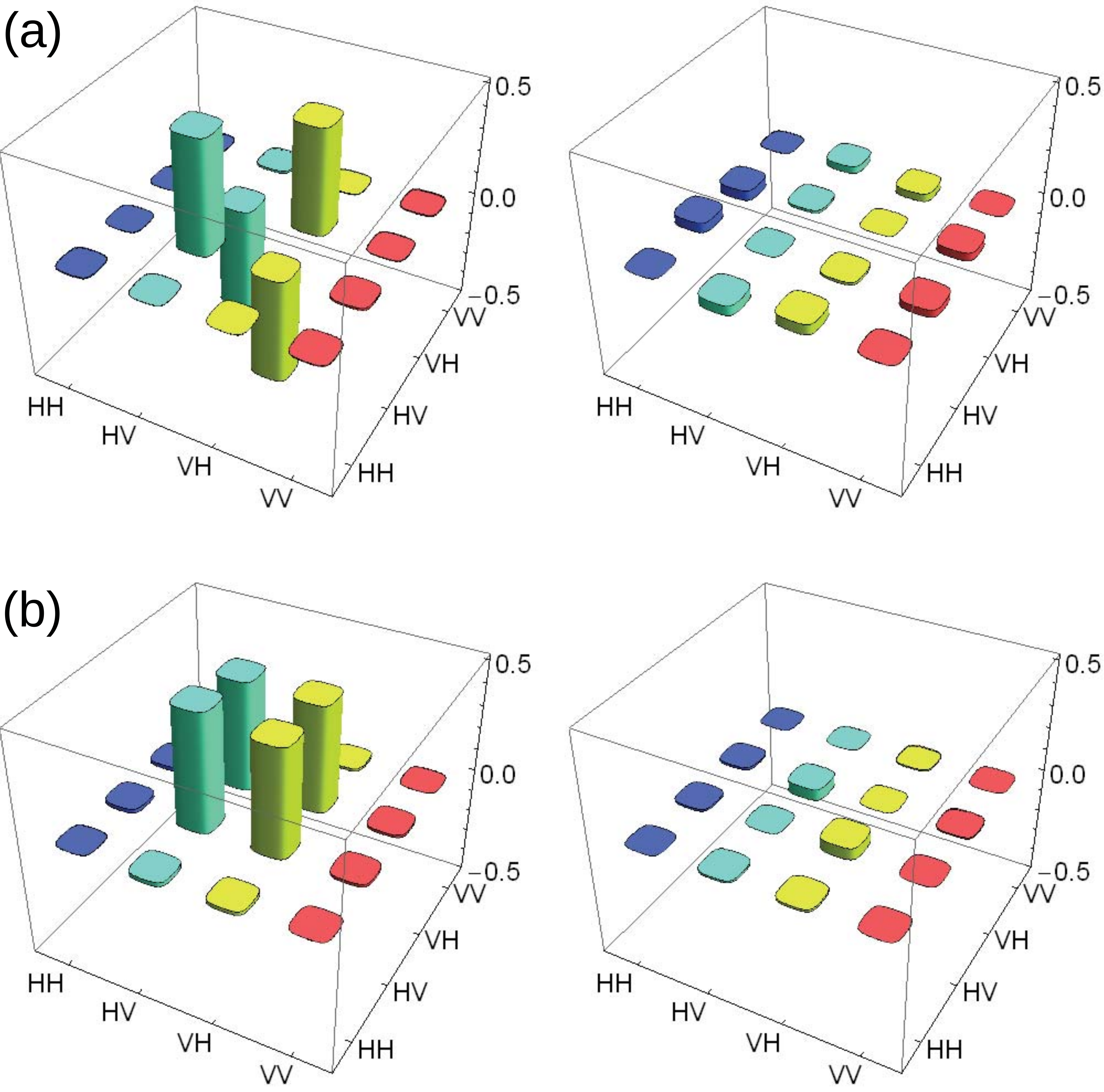}
\caption{ Real (left) and imaginary (right) parts of the reconstructed density matrix. (a) for $\left| {\psi ^ - } \right\rangle$ state, (b) for $\left| {\psi ^ + } \right\rangle$ state.  } \label{DensityMatrix}
\end{figure}

We also carried out state tomography  of our two-photon polarization state.
Polarizer 1 and  2  in Fig. \ref{setup} were replaced by  combinations of HWP, QWP and PBS, to allow polarization correlation analysis in not only linear but also circular polarization bases.
The density matrix $\rho_{exp}$ reconstructed with a maximum likelihood estimation method \cite{James2001} is shown in  Fig. \ref{DensityMatrix}.
The  fidelities \cite{Jozsa1994}, $F \equiv \left\langle {\psi ^ \pm} \right|\rho _{\exp } \left| {\psi ^ \pm  } \right\rangle $, to the ideal Bell state $|\psi^\pm \rangle$, were estimated as 0.981 $\pm$ 0.0002 (0.973 $\pm$ 0.0002) for $\left| {\psi ^ -  } \right\rangle$ and 0.980 $\pm$ 0.001  (0.968 $\pm$ 0.001) for $\left| {\psi ^ +  } \right\rangle$ with  background subtracted data (raw data) accumulated in 10s.
The corresponding concurrences \cite{Wootters1998}  were 0.981 $\pm$ 0.0004 (0.971 $\pm$ 0.0008) and  0.969 $\pm$ 0.002 (0.956 $\pm$ 0.002), respectively.
%
%
These values indicated that our  states were highly entangled.

\begin{figure}[tbp]
\centering\includegraphics[width=6cm]{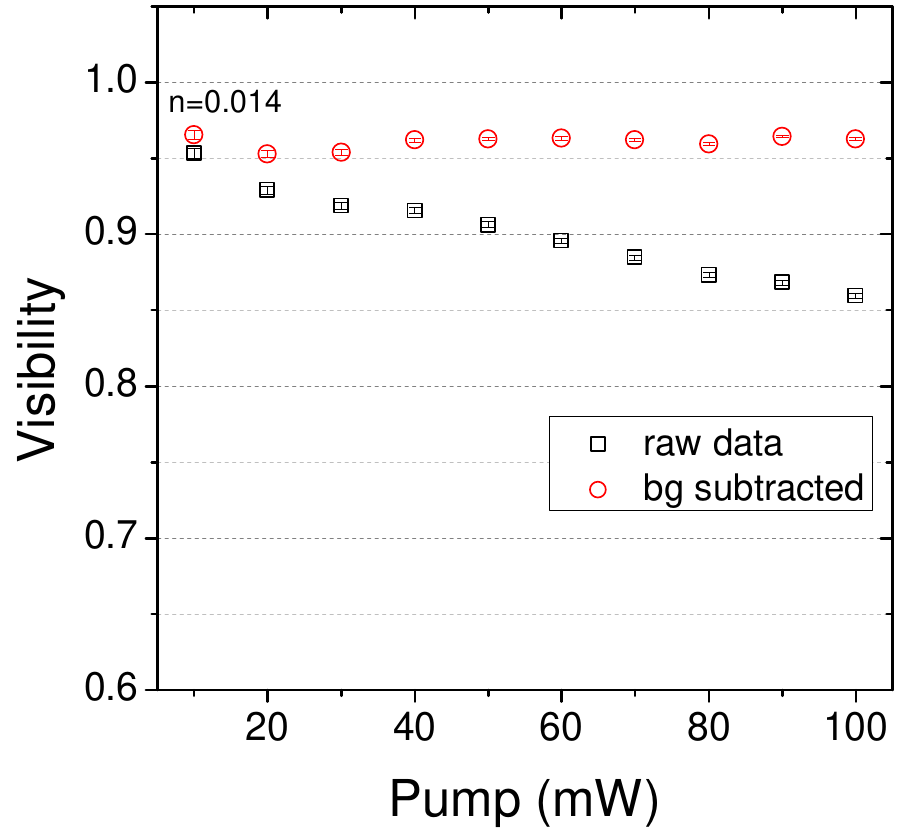}
\caption{ Raw and background subtracted  visibilities with Polarizer 1 set at  45 degrees  for  the $\left| {\psi ^ - } \right\rangle$ state as a function of incident pump power.
 The uncertainties of these visibilities were derived using Poissonian errors on the coincidence counts. The left two points corresponds to the data in Fig. 3 and 4, with an average photon numbers per pulse of 0.014.
 } \label{VandPump}
\end{figure}

To investigate the effect of multi-pair emission on the entangled state, we measured the visibility  as a function of pump power, as shown in Fig. \ref{VandPump}.
The raw visibilities exhibited a linear decrease with respect to the increase of the pump power, providing
evidence of multi-pair generation at higher pump power \cite{Evans2010}.
After background subtraction, the visibility was almost fixed at 96\% for pump power from 10 mW to 100 mW.
In the future, this low visibility at high pump power can be improved by using a laser with high repetition rate and low average power per pulse, such as the  10 GHz-repetition-rate comb laser \cite{Morohashi2008}.

\section{Discussion and Outlook}

Comparing our scheme with the previous entangled photon source with a GVM-PPKTP  crystal in a calcite beam displacer configuration  at telecom wavelength \cite{Evans2010},
our count  rates are more than 100 times higher, mainly thanks to our highly efficient SNSPDs and fine alignment in Sagnac-loop.
Obtaining entangled photon pairs with a high count  rate  and a  low pump power at telecom wavelength is an important feature of our scheme.

In this experiment, we set the wavelength of the signal and the idler to be the same.
By changing the temperature, we could also obtain non-degenerated entangled photons, which can be used to prepare a frequency-entangled state or hyper-entangled state at telecom wavelengths \cite{Barreiro2005}.
Since the PPKTP crystal has the property of a spectrally wide  tunablility with high purity \cite{Jin2013OE}, the source can also be a wavelength-widely-tunable entangled photon source   by using broadband dual-wavelength HWP and PBS in Fig. \ref{setup}.

In the future, this polarization-entangled photon source will be useful for a  variety of applications in quantum information and communication at telecom wavelengths.
For example, this  source is directly applicable to free space quantum key distribution at telecom wavelengths and short distance fiber communications.
It can also  be applied to quantum communication experiments using multiple entangled photon sources, such as quantum teleportation and entanglement swapping.
By changing this polarization-entangled photon source to a time-bin entangled photon source by simply using a Mach-Zehnder-type delay system, this source can be used for a long-distance fiber-based quantum key  distribution system.

\section{Conclusion}
In summary, we have demonstrated a polarization-entangled photon source with a PPKTP crystal in a Sagnac interferometer configuration.
The PPKTP crystal satisfies the GVM condition at telecom wavelengths, therefore, the intrinsic spectral purity is much higher than that at the near-infrared wavelength range.
We have achieved visibilities of over 96\% in quantum correlation measurement, an $S$ value of 2.76 $\pm$ 0.001 in Bell's inequality measurement, and fidelities of  0.98 $\pm$ 0.0002 in quantum state tomography.
The photons were detected by highly efficient SNSPDs and  coincidence counts of 20 kcps were achieved at 10 mW pump.
This entangled photon source is compact, robust, highly-entangled, ultra-bright and spectrally highly pure.
Our GVM-PPKTP-Sagnac scheme will be useful for  quantum information and communication systems.

\section*{Acknowledgments}
The authors are grateful to N. Matsuda, N. Singh, F. Wong and R. Ursin for helpful discussions.
This work was supported by the Founding Program for World-Leading Innovative R\&D on Science and Technology (FIRST).

\end{document}